\documentclass[twocolumn, superscriptaddress,aps,prl]{revtex4}%
\usepackage{amsfonts}
\usepackage{amsmath}
\usepackage{amssymb}
\usepackage{graphicx}%
\begin{document}
\title{Non-Markovian Dynamics Impact on the Foundations of Statistical Mechanics }
\author{Heng-Na Xiong}
\affiliation{Department of Physics and Center for Quantum Information Science, National Cheng Kung University, Tainan 70101, Taiwan}
\author{Ping-Yuan Lo}
\affiliation{Department of Physics and Center for Quantum Information Science, National Cheng Kung University, Tainan 70101, Taiwan}
\author{Wei-Min Zhang}
\email{wzhang@mail.ncku.edu.tw}
\affiliation{Department of Physics and Center for Quantum Information Science, National Cheng Kung University, Tainan 70101, Taiwan}
\date{Oct. 15, 2013}
\author{Franco Nori} 
\affiliation{Center for Emergent Matter Science, RIKEN, Saitama 351-0198, Japan}
\affiliation{Physics Department, The University of Michigan, Ann Arbor, Michigan, 48109-1040, USA}
\author{Da Hsuan Feng}
\affiliation{Department of Physics, National Tsing Hua University, Hsinchu 30013, Taiwan}

\begin{abstract}
The foundations of statistical mechanics, namely how equilibrium hypothesis emerges microscopically from quantum theory, is explored through investigating the environment-induced quantum decoherence processes. Based on the recent results on non-Markovian dynamics [Phys. Rev. Lett. 109, 170402 (2012)], we find that decoherence of quantum states manifests unexpected complexities. Indeed, 
an arbitrary given initial quantum state, under the influence of different reservoirs, can evolve into four different steady states: thermal, thermal-like, quantum memory and oscillating quantum memory states. The first two steady states \textit{de facto} provided a rigorous proof how the system relaxes to thermal equilibrium with its environment. The latter two steady states, with strong non-Markovian effects, will maintain the initial state information and not reach thermal equilibrium, which is beyond the conventional wisdom of statistical mechanics.  
\end{abstract}
\maketitle

Statistical mechanics were established on the equilibrium hypothesis that over a sufficiently long time a given system can always reach thermal equilibrium with its environment, and  the statistical distribution does not depend on its initial state \cite{Landau69,Kubo85}. For over a century, investigating the foundations of statistical mechanics has been focused on the questions \cite{Huang87}: (i) how does macroscopic irreversibility emerge from microscopic reversibility? and (ii) how does the system relax to thermal equilibrium with its environment? Obviously, the foundations of statistical mechanics and the answers to these questions rely on a deep understanding of the dynamics of  systems interacting with their environments. Recent results on environment-induced non-Markovian decoherence dynamics \cite{Zhang12} enable us to address these fundamental problems. 
Decoherence is also a main concern in the recent developments of quantum information technology \cite{Wineland00,Haroche08}. 
The current understanding of decoherence dynamics has provided answers to several fundamental issues, such as quantum measurement and the quantum-to-classical transition \cite{Leggett87,Zurek03}. In the past two decades, many theoretical and experimental investigations were devoted to this topic, most of these taking the memory-less (Markov) limit \cite{Breuer02, Weiss08}. However, experimental implementations of nanoscale solid-state quantum 
information processing \cite{Liu10,Xiang13} makes strong non-Markovian memory effects unavoidable, thus rendering their study a pressing and vital issue \cite{An07,Paz08,Wolf08,Tu08,Zhang12,Breuer09,Xiong10,Kossakowski10,Znidaric11,Lei11,Liu11,Madsen11}. 

As it is well-known, any realistic quantum system inevitably interacts with its environment. When such interaction is not negligible, the system must be treated as an open system. 
Understanding the dynamics of open systems is one of the most challenging topics in physics, chemistry, and biology.   
The dynamical evolution of an open quantum system is determined by the master equation. 
The master equation plays the same role for open quantum systems as the Newtonian equation for macroscopic objects, the Maxwell equation for electromagnetic fields, and the Schr\"{o}dinger equation for isolated quantum systems. In 1928, the first master equation was phenomenologically introduced by Pauli \cite{Pauli28}. Since then, many progresses were made with various approaches in deriving the master equation for various different open quantum systems \cite{Weiss08,Breuer02}, in particular, for quantum Brownian motions \cite{Feynman63,Leggett83,Hu92}. However, there is still a lack of satisfactory answers.  Indeed, not having a rigorous and solvable master equation remains a primary obstacle to understand the foundations of statistical mechanics. 

Recently, 
we have derived microscopically the exact master equation \cite{Tu08,Jin10,Lei12} for noninteracting fermion (boson) systems coupled,  via particle-particle exchanges, to various noninteracting fermion (boson) reservoirs, using the coherent state path-integral formulation \cite{Zhang90} and the Feynman-Vernon influence functional method \cite{Feynman63, Weiss08}. The result is
\begin{align}
\dot{\rho}(t) & = -i\big[  H'_{S}(t)
,\rho(t)\big]  \nonumber\\
& +\sum_{ij}\big\{\gamma_{ij}\left(  t\right)
\big[2a_{j}\rho(t) a_{i}^{\dagger} -a_{i}^{\dagger}a_{j}\rho(t)  -\rho(t)
a_{i}^{\dagger}a_{j}\big] \nonumber\\
&+\widetilde{\gamma}_{ij}(t) \big [a_{i}%
^{\dagger}\rho(t) a_{j}  \pm a_{j}\rho(t)  a_{i}^{\dagger}-a_{j}^{\dagger}a_{i}%
\rho(t)  \mp\rho(t) a_{j}a_{i}^{\dagger}\big]\big\} .
\label{Exact-ME}%
\end{align} Here $\rho(t)$ is the reduced density matrix describing the state of the system;
$H'_S(t)$ is the renormalized system Hamiltonian after the environmental degrees of freedom are completely integrated out; $a_{i}^{\dagger}$ ($a_{i}$) is the particle creation (annihilation) operator; the up (down) sign of  $\pm$ or
$\mp$ in the third line, and also hereafter, corresponds to the system being bosonic (fermionic). The environment-induced energy dissipation (i.e.,~relaxation) coefficients $\gamma_{ij}(t)$ and thermal fluctuations (i.e.,~noises) coefficients $\widetilde{\gamma}_{ij}(t)$ are determined microscopically and exactly from Keldysh's nonequilibrium Green functions \cite{Jin10,Lei12,supm}. These single-particle nonequilibrium Green functions obey Dyson equations that include nonperturbatively all environment-induced non-Markovian memory effects. Thus, the exact master equation showed that single-particle dynamics in an open system is intrinsically irreversible \cite{Zhang12}, so is the system. The remaining question for the foundations of statistical mechanics could be answered by solving the exact master equation. Solving this, one could monitor the evolution of an open quantum system, and its steady state, thus unambiguously unveiling the microscopic statistical process.
\begin{figure}[ptb]
\centering
\includegraphics[width=0.48\textwidth]{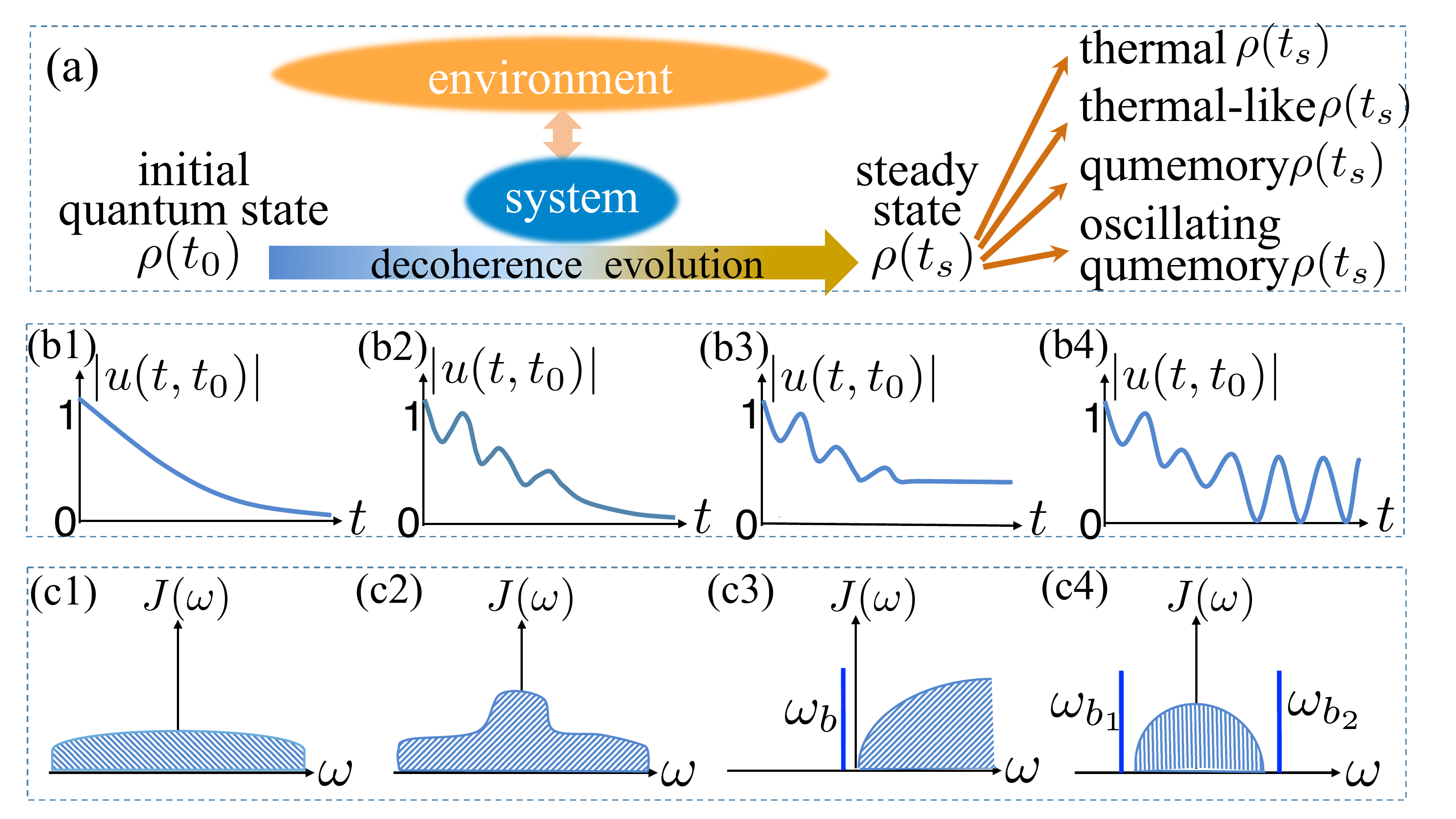}\newline\caption{
(a)  An arbitrary initial quantum state $\rho(t_0)$ of the system, under the influence of different environments, can decohere into one of four possible steady states $\rho(t_s)$.
(b1-b4) The four decoherence scenarios originate from four different relaxation processes, described by the dissipative propagating function $u(t,t_0)$, upon the non-Markovian effects being weak or strong.  (c1-c4) Weak or strong non-Markovian effects rely on the different spectral density structure and also on the strength of the system-environment couplings, given by $J(\omega)$.}
\label{fig-QCT}%
\end{figure}

An open system is usually defined as a system consisting of only one or a few relevant dynamical variables in contact with a huge environment \cite{Weiss08}. To be specific, here we consider a system with one
dynamical variable coupled to a general environment having infinite dynamical variables.  The environment is specified by the spectral density $J(\omega)$  and is initially in a thermal equilibrium state $\rho_E(t_0)$  at temperature  $T$, with the particle distribution $\overline{n}(\epsilon,T) \!=\! 1/[e^{(\epsilon-\mu)/k_BT} \mp 1]$. The system initially is in an arbitrary pure quantum state:  $|\psi(t_0)\rangle \!=\! \sum_{n=0}^\infty c_n |n\rangle$ ($n$ only takes $0$ and $1$ for fermion systems), with density matrix
\begin{align}
\rho (  t_{0})  =\sum_{m,n}c^*_{m}c_{n}|m\rangle\langle n|  . \label{rho-is}
\end{align}
Here the diagonal matrix element $|c_n|^2$  represents the probability for the system in the particle-number state $|n\rangle$ so that $\sum_n |c_n|^2 =1$. The off-diagonal matrix element  $c^*_mc_n$  ($n\neq m$) represents the quantum coherence between  $|n\rangle$ and $|m\rangle$. The initial state of the total system, $\rho_{\rm tot}(t_0)=\rho(t_0)\otimes \rho_E(t_0)$, will follow a unitary (reversible) evolution. However, the system and the environment are initially not in equilibrium, and the evolution of the system undergoes a nonequilibrium process, described by the reduced density matrix  $\rho(t)$. Solving the master equation with the general initial state (\ref{rho-is}), we find that 
\begin{equation}
\rho\left(  t\right)  =\! \! \sum_{m,n}c^*_mc_n \!\!\!\! \sum_{k=0}^{\min
\{m,n\}}\!\!\!\!  d_{k}\mathcal{A}_{mk}^{\dagger}\left(  t\right)  \widetilde{\rho
}(t )  \mathcal{A}_{nk}\left(  t\right)  .
\label{rho-t-general}
\end{equation}
Here the kernel $\widetilde{\rho}( t) \!=\!\sum_{n'}\frac{
 [v(t,t) ]^{n'}}{[ 1\pm v(t,t) ]  ^{n'\pm 1}}|n'\rangle\langle
n'|$, the operator $\mathcal{A}_{mk}^{\dagger}(t)
\!=\!\frac{\sqrt{m!}}{( m-k) !\sqrt{k!}}\big[\frac{u(t,t_0)}{1\pm v(t,t) }
a^{\dagger}\big]^{m-k}$, and the coefficient $d_{k}\!=\!\big[ 1-\frac{|u(t,t_0)|
^{2}}{1\pm v(t,t)}\big]^{k}$.  The solution (\ref{rho-t-general})  shows that $\rho(t)$  is fully determined by the  single-particle dissipative propagating function $u(t,t_0)$  and correlation function  $v(t', t)$ with the relation $\langle a^\dag(t)a(t')\rangle=u(t',t_0)\overline{n}(t_0)u^*(t,t_0)+v(t',t)$, where $t'\leq t$, and $\overline{n}(t_0)$ is the initial average particle number in the system. As it has been shown \cite{Tu08,Jin10,Lei12}, these two Green functions (corresponding to the retarded and correlation Green functions in Keldysh's formalism \cite{Keldysh65}) are uniquely determined by the spectral density  $J(\omega)$, which coincides with the conclusion reached by Leggett \textit{et al.} \cite{Leggett87} that for any problem in which a thermal equilibrium statistical average is taken over the initial states of the environment and a sum over the final states, complete information about the effect of the environment is encapsulated in the single spectral density.

The spectral density $J(\omega)$ is microscopically defined as a multiplication of the environmental density of states with the system-environment coupling strength. Due to different degrees of non-Markovian memory, different spectral densities can induce vastly different dissipations and fluctuations \cite{Zhang12}. Indeed, this gives rise to four decoherence scenarios depicted in Fig.~\ref{fig-QCT}(a): A given arbitrary initial quantum state $\rho(t_0)$ of the system can decohere into one of four possible steady states $\rho(t_s)$: thermal state, thermal-like state, qumemory (quantum memory) state, and oscillating qumemory state, here $t_s$ denotes the time scale the system reaching its steady state.  The former two provide indeed a rigorous foundation for statistical mechanics. They showed how a system is relaxed into equilibrium with its environment.  The latter two go beyond standard statistical mechanics, namely the system would memorize its initial state information and not reach equilibrium. Which scenario will the system ultimately evolve into is determined by the dissipation dynamics. Thermal noise, given by fluctuating correlations, is naturally manifested through the fluctuation-dissipation relation. 
Below we shall detail these four decoherence scenarios. 

The recent investigation \cite{Zhang12} show that the general non-Markovian dynamics of open quantum systems consists of nonexponential dissipative dampings and dissipationless oscillations. When the non-Markovian memory effects are negligible, non-exponential dampings are reduced to the exponential one.
We consider first a system undergoing an exponential damping. Correspondingly, the dissipative propagation function $u(t_s,t_0) \rightarrow 0$, see Fig.~\ref{fig-QCT}(b1). Then the general solution (\ref{rho-t-general}) will evolve into a steady state of the form
\begin{align}
\rho\left(  t_{s}\right)\!
 \rightarrow \!\left\{ \begin{array}{ll}\! \sum_{n=0}^{\infty}\frac{[v(t_s,t_s) ]  ^{n}} 
 {[  1+v(t_s,t_s) ]  ^{n+1}}|n\rangle\langle n|  & {\rm for~bosons} \\ \!\! \big[1-v(t_s,t_s) \big]|0\rangle \langle 0|+ v(t_s,t_s)|1\rangle \langle 1| & {\rm for~fermions} \end{array} \right. .
\label{rho-ts-1}%
\end{align}
Here $v(t_s,t_s)$ is the steady-state value of the single-particle correlation (fluctuation) function, which is found \cite{Xiong10,supm} in the exponential damping:  $v(t_s,t_s)\!=\!\overline{n}( \epsilon_S,T) \!=\!1/[e^{\epsilon_S-\mu)/k_BT}\! \mp \!1]$, where $\epsilon_S$ is the single particle energy in the system, and $T$ is the environment temperature. This process usually occurs when  system-environment couplings are sufficiently weak (compared with the characteristic energy scale of the system) and/or the spectral density has a rather wide distribution, see Fig.~\ref{fig-QCT}(c1). Consequently,  particle correlations in the environment have nearly a delta-function profile in the time domain, and the system will loss all memory during its evolution. Ultimately, the initial quantum coherence is washed out and the steady state is independent of the initial state of the system. In other words, the system finally reaches thermal equilibrium with its environment. This is precisely given by the solution (\ref{rho-ts-1}), which is the standard \textit{thermal state}. It should be underscored that even if we had not made the assumption of large particle number in the system, this solution already provided the microscopic answer to the foundations of statistical mechanics: the system is eventually thermal equilibrium with its environment if there is no driving field acting on it. 

For many real open systems, however, their dynamics do not follow an exponential damping. This is the case when the system-environment couplings become stronger or the reservoir has band structures. The corresponding dissipation dynamics usually manifests a nonexponential decay and sometimes even becomes dissipationless \cite{Zhang12}. Therefore, the quantum state may not evolve into the expected thermal state.  Specifically, we shall now consider the spectral density to be slightly different from a wide distribution (e.g.~Fig.~\ref{fig-QCT}(c2)) or it may be very different from a wide distribution (e.g.~Fig.~\ref{fig-QCT}(c3-c4)) but has a weak system-environment coupling. In these cases, in a short time scale the system retains partial memory on the information exchange with the environment. Consequently, the dissipation process exhibits a short-time oscillation over the exponential damping profile, although ultimately $u(t_s,t_0) \rightarrow 0$, see Fig.~\ref{fig-QCT}(b2). This manifests a weak (non-Markovian) memory effect. The system will reach a steady state with the same form as Eq.~(\ref{rho-ts-1}), 
but the particle correlation function $v(t_s,t_s)=\int d\epsilon \mathcal{D}( \epsilon) \overline{n}(  \epsilon,T) \neq \overline{n}(\epsilon_S,T)$, where $\mathcal{D}( \epsilon)$ is the environment-modified density of states of the system. Since the steady-state particle distribution is different from the simple thermal particle distribution, the corresponding steady state shall be referred to as a \textit{thermal-like state}. Notice that the memory effect in this scenario exists only in a quite short time, so that the steady-state solution is also independent of the initial state of the system. The quantum coherence at the end is also lost. The thermal equilibrium hypothesis of statistical mechanics remains intact. In other words, as long as the particle propagation in the system is a continuous damping under weak non-Markovian memory, the system and environment will eventually reach thermal equilibrium. 

It is especially interesting when the spectral density does not cover the entire frequency regime, as shown in Fig.~\ref{fig-QCT}(c3), and also has a strong system-environment coupling. In this case, the strong system-environment coupling could induce a localized state at $\omega_{b}$. The dissipation dynamics, shown in Fig.~\ref{fig-QCT}(b3), is significantly different from the previous two cases. In particular, the decay process even ceases after some time and the system becomes dissipationless with $|u(t_s,t_0)|$  approaching a nonzero constant. This is due to strong non-Markovian memory \cite{Zhang12}. The quantum state of the system will then evolve into a steady state that bears no resemblance to that of (\ref{rho-ts-1}). It keeps the same form as the solution (\ref{rho-t-general})
\begin{equation}
\rho\left(  t_s\right)  =\!\!\! \sum_{m,n}c^*_mc_n\!\!\!\! \sum_{k=0}^{\min
\{m,n\}}\!\!\!\!\!  d_{k}{\mathcal{A}}_{mk}^{\dagger}\left(  t_s\right)  \widetilde{\rho
}(t_s)  {\mathcal{A}}_{nk}\left(  t_s\right)  .  \label{rho-ts-3}
\end{equation}
with $u(t_s,t_0)\rightarrow A\exp\{-i\omega_b(t_s-t_0)\}$,
and $A$ is the particle field amplitude arising from the localized state. 
The particle correlation function is given by $v(t_s,t_s)=\int \!\!d\epsilon[\mathcal{D}_b(\epsilon,t_s)+\mathcal{D}(\epsilon)]
\overline{n}( \epsilon,T ) $, where $\mathcal{D}_b(\epsilon,t_s)$ emerges from the localized state. The solution (\ref{rho-ts-3}) is no longer a thermal or thermal-like state. The off-diagonal matrix elements in the state (\ref{rho-ts-3}) do not vanish, thus signaling the maintenance of some quantum coherence. Furthermore, the average particle number in (\ref{rho-ts-3}) is given by  $\overline{n} (  t_s )  =\left\vert u ( t_s,t_0 )\right\vert^{2}n (  t_{0})  +v( t_s,t_s )$, which explicitly depends on the initial average particle number $\overline{n}(t_0)$ in the system. In this scenario, the system has a very long memory and does not reach the thermal equilibrium with its environment. This property remains valid when the system contains a macroscopically large particle number, and therefore is not expected from the conventional wisdom that thermal equilibrium can always be reached with its environment \cite{Landau69,Huang87}. 
In this case, through non-Markovian memory processes, the system remains in a stationary but nonequilibirum state, which constantly keeps information (coherence) exchange with its environment. We call this state (\ref{rho-ts-3}) a quantum memory state, or simply a \textit{qumemory state}. 

When the spectral density has a band structure, as shown in Fig.~\ref{fig-QCT}(c4), the system could involve more than one localized state. These localized states produce dissipationless dynamics for the system with different oscillation frequencies. As a result, the system evolves into a steady state having the same form as (\ref{rho-ts-3}) with vastly different dissipationless dynamics. It combines different localized states with $u(t_s,t_0)  =\sum
_{j}A_{b_{j}}\exp\{-i\omega_{b_{j}}(t_s-t_0)\}$,  and forms an \textit{oscillating qumemory state}. As a result, the average particle number,  $\overline{n}(t_s)  =| u(t_s,t_0) |^{2}\overline{n}(  t_{0})  +v(t_s,t_s)  $, not only depends on its initial value but also oscillates in time. This oscillation is purely quantum in nature and can have interesting features for quantum device systems and also implications for quantum information processing against decoherence. It is noteworthy that the localized states have already incorporated all environmental back-action effects and are decoherence-free. In addition, the steady-state oscillations indicate that the system can maintain quantum coherence in a nonequilibirum manner with its environment. This  is also unexpected. 

\begin{figure}[ptb]
\centering
\includegraphics[width=0.48\textwidth]{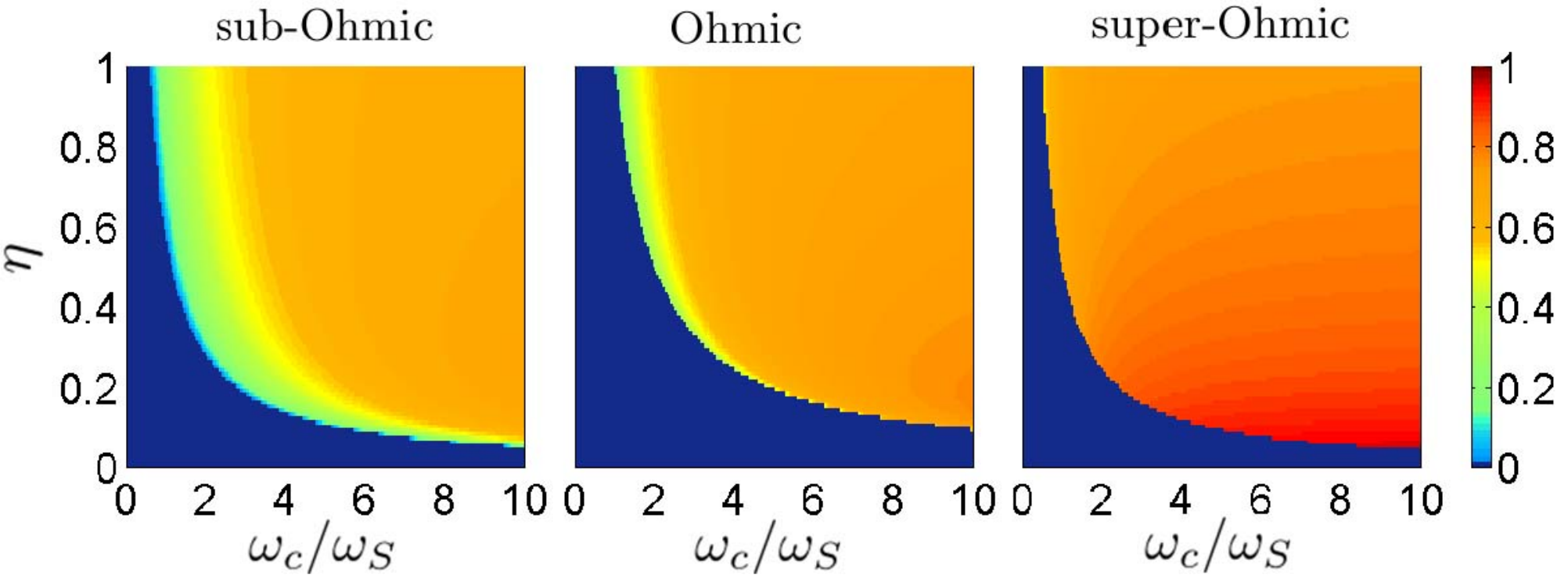}\newline\caption{
Steady values of the single-particle propagating field amplitude  
$|u(t_s, t_0)|$, with sub-Ohmic ($s=1/2$), Ohmic
($s=1$), and super-Ohmic ($s=3$) spectral densities, respectively, versus the
coupling strength $\eta$ and the cutoff frequency $\omega_{c}$.
Here $\omega_S=\epsilon_S$ is the frequency of the system.
For each spectral density, $|u(t_s, t_0)|
 $ remarkably transits from zero to nonzero across the boundary line
$\eta_{c}(\omega_c)=\omega_S/[\omega_{c}\Gamma\left(  s\right) ]$,
which corresponds to the transition from thermal-like states to
qumemory states.}
\label{fig-uts}
\end{figure}

We will now demonstrate how the above four decoherence scenarios can be realized in practice. Since the steady state of an open system depends closely on the spectral density structure, we consider a general environment with the Ohmic-type spectral density: $J\left(  \omega\right)\!=\!2\pi\eta\omega\big( \omega/\omega_{c}\big)^{s-1}\!\! \exp\left(  -\omega/\omega_{c}\right)$, which was proposed by Leggett \textit{et al.}\cite{Leggett87} to simulate a large class of thermal reservoirs. Here $\eta$  is a dimensionless system-environment coupling strength and $\omega_{c}$  is a cutoff frequency. The spectral density can be classified into three regions: sub-Ohmic ($0\!\!<\!\!s\!\!<\!\!1$), Ohmic ($s\!\!=\!\!1$), and super-Ohmic ($s\!\!>\!\!1$). Each region portrays a different environment. The Ohmic-type spectral density takes the structure shown by Fig.~\ref{fig-QCT}(c3) with $\omega \!\!\geq \!\!0$. In this case, there is a localized state with  $\omega_b\!\! <\!\!0$, when the coupling strength $\eta$  exceeds the critical boundary $\eta_{c}(\omega_c)\!\!=\!\!\omega_S/[\omega_{c}\Gamma\left(  s\right) ]$. As shown in Fig.~\ref{fig-uts}, the boundary $\eta_{c}(\omega_c)$ distinguishes the dissipation and dissipationless regimes (through the zero and nonzero values of the single-particle propagating field amplitude $|u(t_s, t_0)|$). This would indicate how and when a thermal-like state or a qumemory state emerges through varying the spectral density. Note that the thermal state (\ref{rho-ts-1}), with $v(t_s,t_s)\!\!=\!\!\overline{n}(\epsilon_S,T)$, corresponds to the limiting case that the height of the spectral distribution $\eta\omega_{c}/\omega_{S}\ll1$  and the spectral width  $\omega_{c}/\omega_{S}\gg1$, where the non-Markovian memory is negligible.

To demonstrate the existence of oscillating qumemory states,
a nanocavity subject to a structured reservoir consisting of coupled 
resonators is considered. By modeling the reservoir as a tight-binding 
1D system, its dispersion takes the form $\omega_{k}
=\omega_{c}-2\xi\cos\left(  2k\right)  $. This leads to the finite-band 
structure of the spectral density 
$J(\omega) =\eta^{2}\sqrt{4\xi^{2}-(\omega-\omega_{c})^{2}}$, 
with band $\vert \omega-\omega_{c}\vert \leq2\xi$, where
$\omega_{c}$ is the band center of the reservoir, $\xi$ is the
intercavity coupling in the reservoir, and $\eta$ is a ratio between the 
cavity-reservoir coupling and intercavity coupling $\xi$. It is easy to 
verify that when the coupling strength $\eta>\eta_{c}=\sqrt{2+|
\Delta|/\xi}$, where $\Delta=\omega_{S}-\omega_{c}$ is the detuning,
there are two localized states $\omega_{b_{\pm}}$ outside of the band \cite{Lei12}. 
As shown in Fig.~\ref{fig-QCT}(c4), these two simultaneously occurring modes 
will cause the dissipation dynamics to oscillate. In particular, in the zero detuning 
case the steady value of the single-particle propagating function oscillates as a cosine function: 
$u\left(  t_{s},t_0\right)  =\frac{\eta^{2}-2}{\eta^{2}-1}e^{-i\omega_{c}t_s}
\cos\big(  \frac{\eta^{2}\xi}{\sqrt{\eta^{2}-1}} t_s\big)$.
This gives rise to an oscillating qumemory state that could be easily 
discovered, since materials with band structures are common in nano-systems.

\begin{figure}[ptb]
\centering
\includegraphics[width=0.48\textwidth]{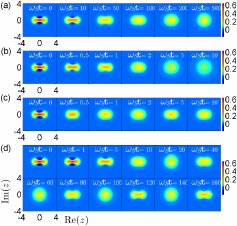}\newline\caption{
Time evolutions of  Schr\"{o}dinger's cat-like states are given for four decoherence scenarios, in terms of contour plots of  Wigner distribution function for (a) Ohmic spectral density with cutoff frequency $\omega_{c}=10\omega_{S}$ and coupling strength
$\eta=0.001$,  where the non-Markovian effect is almost negligible; (b) Ohmic spectral density with  
$\omega_{c}=5\omega_{S}$ and $\eta=0.1<\eta_{c}(\omega_c)=0.2$, 
which is in the weak non-Markovian regime; (c) Ohmic spectral density with  
$\omega_{c}=5\omega_{S}$ and $\eta=0.5>\eta_{c}(\omega_c)$, the strong non-Markovian regime;  (d) the spectral density of a tight-binding 1D system with $\eta=3.0>\eta_{c}=\sqrt{2}$ and band center
$\omega_{c}=\omega_{S}$, also in the strong non-Markovian regime.  Here the initial environment temperature is set as $T=2\omega_{S}$, and the initial coherent amplitude $\alpha_{0}=1$.}
\label{fig-FV-WF}
\end{figure}

The possible observation of the above four decoherence scenarios can be achieved through the decoherence evolution of  Schr\"{o}dinger's cat-like states. Schr\"{o}dinger's cat-like states, which is defined as a superposition of two opposite-moving coherent states  $|\pm \alpha_0\rangle$, have been considered as an archetypical example to demonstrate quantum decoherence \cite{M-H96,Wineland00,Haroche08}. Here we examine the dynamics of such a state in a general environment. In Fig.~\ref{fig-FV-WF},  we show the four decoherence scenarios 
through the  time evolution of a cat-like state, in terms of Wigner function. 
It shows that in the Markov (memory-less) process, the initial cat-like state smoothly loses all its quantum coherence and becomes a thermal state, the corresponding Wigner function becomes a thermal Gaussian distribution located at the origin, see the last graph in Fig.~\ref{fig-FV-WF}(a). 
For other non-Markovian processes, the cat-like state takes on different decoherence behaviors. For a weak non-Markovian decay given in Fig.~\ref{fig-FV-WF}(b), although the steady state (reached at ${t}=10/\omega_{S}$) is thermal-like, decoherence occurs much faster than the thermal state (it takes ${t}=500/\omega_{S}$  to reach the steady state as shown in Fig.~\ref{fig-FV-WF}(a)). 
Furthermore, when the strong non-Markovian effect is dominated by the localized mode(s), the Winger function never becomes a thermal Gaussian distribution. It shows a non-thermal  distribution or an oscillating non-thermal distribution pattern, see Fig.~\ref{fig-FV-WF}(c) or \ref{fig-FV-WF}(d), 
which correspond precisely to the qumemory state or the oscillating qumemory state.
The more detailed dynamics (the computer-generated movies) for the time evolutions of the cat-like state is presented in the Supplemental Materials.

The foundations of statistical mechanics are investigated here through a general study of quantum decoherence. Besides providing a clear understanding of quantum decoherence with thermal equilibrium, the memory-induced decoherence scenarios presented here reveal new features for open quantum systems: namely, how systems can maintain quantum coherence under environmental influence in the strong non-Markovian-memory regime. Indeed, we showed that these systems will not inevitably evolve into thermal equilibrium with its reservoir, and can maintain quantum oscillations in  localized states that are decoherence-free.  These results depend only on the environmental structure characterized by the spectral density and therefore are generic. The unexpected new features may not only provide new insights for the development of quantum information processing, but also advance the study of statistical mechanics for open systems when thermal equilibrium is not reachable.

\noindent\textbf{Acknowledgments \ }We acknowledge the support from the National Science Council of ROC under Contract No.  NSC-102-2112-M-006-016-MY3, and the National Center for Theoretical Science of ROC. 

\noindent\textbf{Author Contributions \ }W. M. Zhang proposed the ideas and methods; H. N. Xiong and P. Y. Lo performed the calculations; H. N. Xiong prepared an initial draft; W. M. Zhang and D. H. Feng interpreted the physics and wrote the manuscript; F. Nori participated in the discussions and contributed to the writing of the manuscript.

\end{document}